\documentclass[twocolumn,aps,showpacs,prb,tightenlines,amsmath,amssymb,superscriptaddress]{revtex4}
\usepackage{graphicx}
\usepackage{amssymb}
\usepackage{dcolumn}
\usepackage{amsmath}
\usepackage{bm}

\usepackage{colordvi}

\begin{document}

\title{Robust strongly-modulated transmission of a $T$-shaped structure with
  local Rashba interaction}
\author{K. Shen}
\affiliation{Hefei National Laboratory for Physical Sciences at
Microscale, University of Science and Technology of China, Hefei,
Anhui, 230026, China}
\author{M. W. Wu}
\thanks{Author to  whom correspondence should be addressed}
\email{mwwu@ustc.edu.cn.}
\affiliation{Hefei National Laboratory for Physical Sciences at
Microscale,
University of Science and Technology of China, Hefei,
Anhui, 230026, China}
\affiliation{Department of Physics,
University of Science and Technology of China, Hefei,
Anhui, 230026, China}
\altaffiliation{Mailing address.}

\date{\today}

\begin{abstract}
We propose a scheme of spin transistor using a $T$-shaped structure
with local Rashba interaction. A wide antiresonance energy gap appears
due to the interplay of two types of interference, the Fano-Rashba interference and
the structure interference. A large current from the gap area can be obtained via
changing the Rashba strength and/or the length of the sidearm by using gate voltage.
The robustness of the antiresonance gap against strong disorder is demonstrated and shows the
feasibility of this structure for the real application.
\end{abstract}


\maketitle

Spin-polarized electron transport has attracted much attention
recently due to the promising application of spintronic
devices.\cite{spintronics}
One of such devices, the Datta-Das
spin field effect transistor,\cite{datta} was proposed by utilizing
the spin precession due to the Rashba
effective magnetic field.\cite{rashba} After that, the role of the
Rashba spin-orbit coupling (SOC) in ballistic transport systems has been extensively
studied.\cite{sato} Very recently, the effect of local
Rashba spin-orbit interaction in quasi-one
dimensional quantum wires was investigated.\cite{sanchez2} As
reported, there exists at least one bound state localized in the
vicinity of the Rashba region due to the equivalent attractive
potential from local Rashba interaction. Such bound state can interfere with
direct propagating channels, leading to the Fano asymmetric
lineshapes\cite{fano,chu,nockel} of the
transmission.\cite{sanchez2} Therefore, this effect was called
the Fano-Rashba effect. Similar Fano-type inference effect in the quantum wire with an
applied magnetic field was also reported, which was based on the
interplay of the transmission channel with
certain spin and the evanescent mode with opposite
spin.\cite{sanchez} The transmission zero dip at Fano antiresonance was proposed to
be helpful in realizing spin transistor by
Sanchez {\em et al.} very recently.\cite{sanchez} However, the robustness of this
proposal against the
disorder, which is essential for real application, remains questionable. In
this report, we will show that the occurrence of
such dip is strongly limited by the disorder. We further propose a
scheme of device using $T$-shaped
structure\cite{sols,feng} with local Rashba interaction.
This device can provide a large energy window for antiresonance in contrast to
single energy points in the ordinary antiresonance devises,
with strong robustness against disorder.

\begin{figure}[bth]
\centering
\includegraphics[width=7cm]{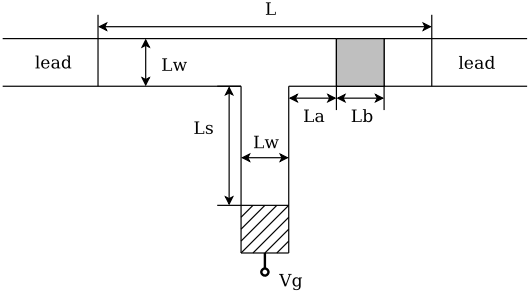}
\caption{ Schematic of $T$-shaped structure with local Rashba
  interaction. The Rashba area is shown as the gray area. The shadowed area
stands for the gate used to tune the length of the sidearm.}
\label{fig1}
\end{figure}

A schematic of the waveguide in our study is shown in Fig.\ \ref{fig1}, where a
waveguide of length $L$ with a sidearm protruding from the center, is
connected to the half-metallic leads through perfect ideal ohmic
contacts. We assume the electron states at the Fermi level are all spin-down ones in the
leads, so that only spin-down electrons can propagate into/out of the
$T$-shaped structure. The effective length of the sidearm $L_s$ can be adjusted
electronically by a gate voltage $V_g$.\cite{feng} The finite width
of the waveguide $L_w$ gives the propagation threshold as the first
quantized subband along the transversal direction.
A perpendicular magnetic field is applied uniformly on the whole device. This field
shifts the energy spectrum by a Zeeman splitting
$V_\sigma=\sigma V_0$.  We neglect the effect of magnetic field on
 orbital motion by assuming the magnetic field is weak and
hence the Landau level is negligible.
The interference of the different Feynman paths
makes it possible to realize spin transistor using $T$-shaped
structure.\cite{sols} However, this kind of transistor is also
strongly limited by the disorder  as we will show later.
In order to get a robust transistor, we further introduce the local Rashba interaction as the gray
area in Fig.\ \ref{fig1}.  From Fano-type interference effect due to the local Rashba
interaction, the transmission of
the propagating channel can also be strongly modulated, especially at the
Fano antiresonance.\cite{sanchez,sanchez2} We
demonstrate that when the individual structure antiresonance and  Fano antiresonance are
close to each other, there exists  a broad  energy window  in which the conductance is zero, i.e.,
there exists an antiresonance energy gap.  This is in contrast to the single (specific) energy
when only the structure or Fano antiresonance is involved.
Moreover, this antiresonance energy gap is very robust against the disorder.
We also find that a large current can be obtained in this gap area
by adjusting the Rashba coupling strength\cite{nitta} and/or the length of the
sidearm using the gate voltage. Such features are very useful for the spin transistors.

We describe the $T$-shaped structure by the tight-banding Hamiltonian
with the nearest-neighbor approximation
\begin{eqnarray}
  \nonumber
H&=&\sum_{l,m,\sigma}\epsilon_{l,m,\sigma}c^+_{l,m,\sigma}c_{l,m,\sigma}-
  t\sum_{l,m,\sigma}(c^+_{l+1,m,\sigma}c_{l,m,\sigma}\\
  &&+c^+_{l,m+1,\sigma}c_{l,m,\sigma}+H.C.)+H_R\ ,
  \label{H}
\end{eqnarray}
with two indices $l$ and $m$ denoting the site coordinates along the $x$
and $y$ axes, respectively. The lattice energy
$\epsilon_{l,m,\sigma}=4t+\sigma V_0$, with the hopping energy
$t=\hbar^2/(2m^\ast a^2)$ and the Zeeman splitting $V_0$. Here
$m^\ast$ and
$a$ stand for the effective mass and lattice constant separately.  The last
term in Eq.\ (\ref{H}) describes the Rashba SOC\cite{rashba,mireles}
\begin{eqnarray}
  \nonumber
  H_R&=&\lambda\sum_{l,m,\sigma,\sigma^\prime}[c^+_{l+1,m,\sigma}c_{l,m,\sigma^\prime}
 (i\sigma_y^{\sigma\sigma^\prime})\\
 &&-c^+_{l,m+1,\sigma}c_{l,m,\sigma^\prime}
 (i\sigma_x^{\sigma\sigma^\prime})
 +H.C.]\ ,
\end{eqnarray}
 in which $\lambda=\alpha/2a$ with $\alpha$ representing the Rashba
 coefficient. The summations $(l,m)$ in $H_R$ are performed only in the gray
 area in Fig.\ \ref{fig1}.

\begin{figure}[bth]
\centering
\includegraphics[width=6.cm]{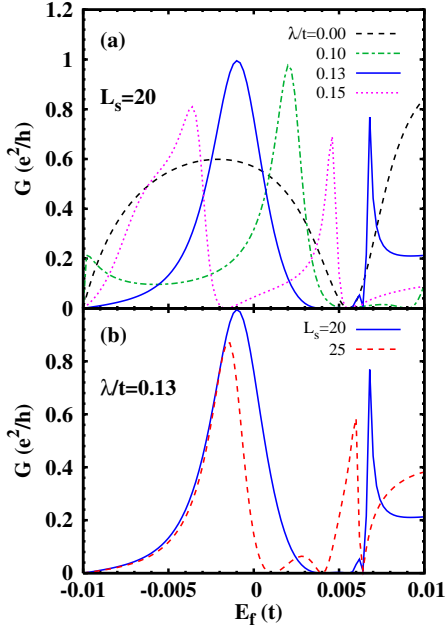}
\caption{(Color online) Conductance {\em vs.} the Fermi energy of the leads (a) with
  different local Rashba SOC strengths when $L_a=L_b=20a$  and (b)
with different sidearm lengths when $L_a=20a$ and $\lambda/t=0.13$.}
\label{fig2}
\end{figure}

The two-terminal conductance is obtained from the Landauer-B\"uttiker
formula\cite{landauer}
\begin{equation}
G^{\sigma\sigma^\prime}(E)=(e^2/h)\mbox{tr}[\Gamma_1^\sigma G_{1N}^{\sigma\sigma^\prime+}(E)
\Gamma_N^{\sigma^\prime}G_{N1}^{\sigma^\prime\sigma-}(E)],
\end{equation}
where $\Gamma_{1(N)}$ denotes the self-energy of the isolated ideal leads and
$G_{1N}^{\sigma\sigma^\prime}$ ($G_{N1}^{\sigma^\prime\sigma}$) is the
retarded (advanced)  Green
function.\cite{datta2}
For energy window $[E,E+\Delta]$, the current is given by $I=\int_E^{E+\Delta}G(\epsilon)d\epsilon$.
We perform a numerical calculation for a waveguide with fixed width
$L_w=20a$. The hard wall potential in the transverse direction gives
the lowest energy of the $n$th subband
$\epsilon_n=2t\{1-\cos[n\pi/(L_w/a+1)]\}$. Throughout this
report, we take the Zeeman splitting energy $V_0=0.01t$ and the Fermi
energy $E_f$ is regard to $\epsilon_1=0.02234t$. The conductance
$G$ represents the conductance of the only propagating spinor $G^{\downarrow\downarrow}$.

In Fig.\ \ref{fig2}(a), the conductance $G$ is plotted as a function of the
Fermi energy at different Rashba coefficients $\lambda$.
In the calculation, $L_s=L_a=L_b=20a$. The result
without the local Rashba SOC is plotted as the
black dashed curve, showing a
transmission zero dip, i.e., the structure antiresonance\cite{feng,shi} dip, at
$E_f\simeq0.005t$. The presence of the local Rashba interaction also
strongly influences the conductance and provides another transmission zero dip,
i.e., the Fano antiresonance dip, as shown by the remaining  curves with $\lambda\not=0$. One finds
that the Fano dip moves to small $E_f$ region with the increase of the SOC
coefficient.\cite{sanchez} The most interesting feature is that
when the Fano dip is close to the structure dip ($\lambda=0.13t$), a wide energy gap [0.0034t,0.0058t]
for antiresonance appears (see the blue solid curve). Moreover, this gap
can be turned off via changing the
SOC coefficient. Specifically, there exists a peak at $\lambda=0.15t$,
which gives a current $I\simeq3\times10^{-4}e^2t/h$ for the energy window
$[0.0040t,0.0045t]$, originally in the gap area. Therefore, it can work as spin transistor
with the on and off features by tuning the Rashba strength with a gate voltage. From
Fig.\ \ref{fig2}(b), one can also see that the on and off features of the transistor can
also be remotely controlled by changing the length of the
sidearm $L_s$ by another gate voltage.\cite{feng}

\begin{figure}[bth]
\centering
\includegraphics[height=4.5cm,width=7cm]{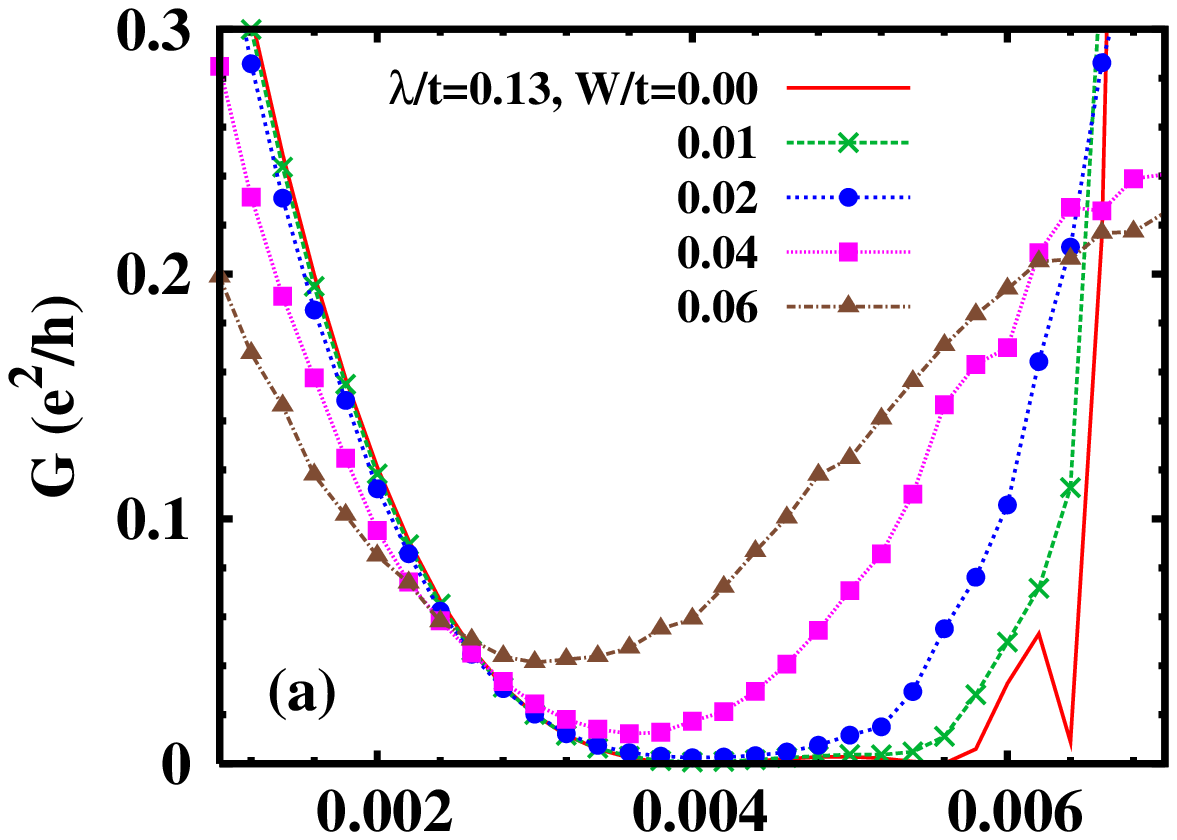}
\includegraphics[height=4.5cm,width=7cm]{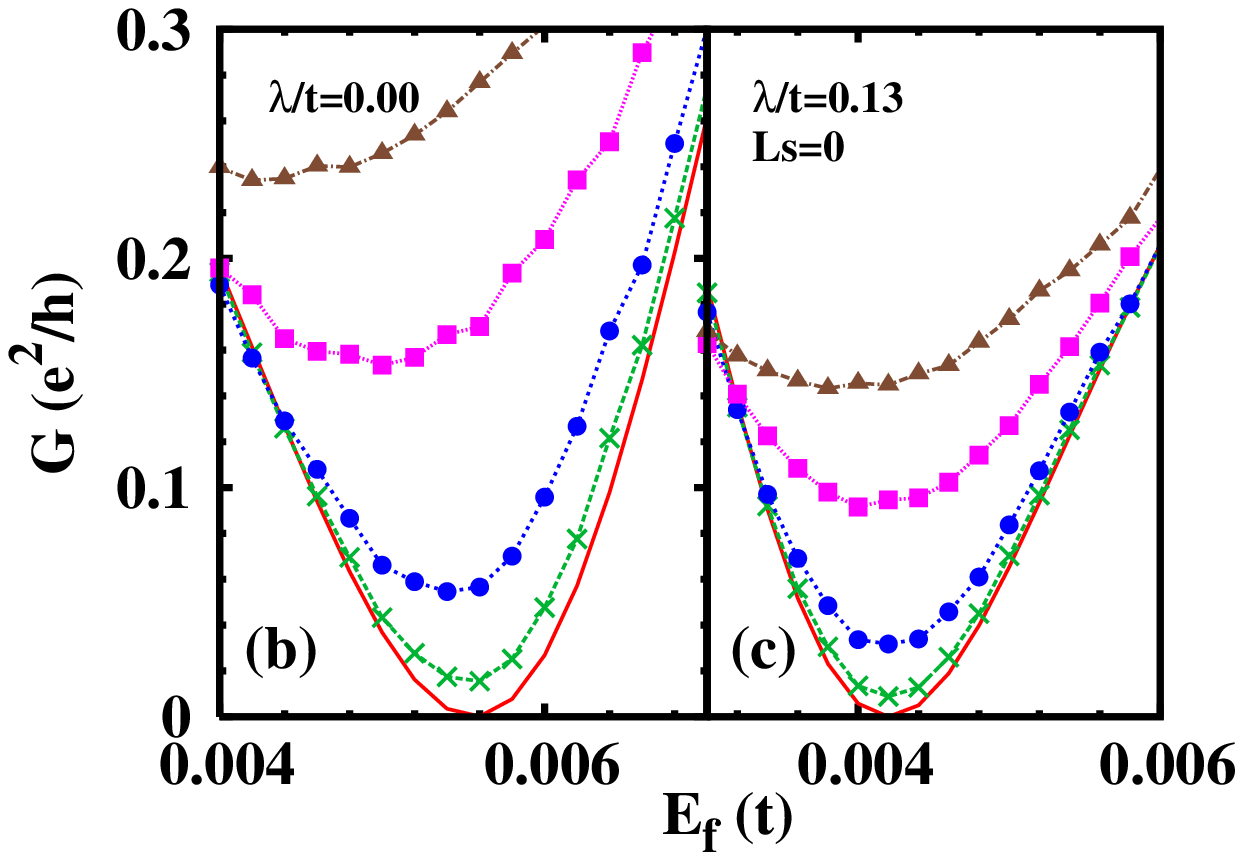}
\caption{(Color online) Conductance vs Fermi energy with different
  disorder strengths $W$ in the vicinity of the antiresonance energy gap. (a) with both the Fano antiresonance and
  the structure antiresonance;
(b) with only the structure antiresonance; (c) with only the Fano antiresonance.}
\label{fig3}
\end{figure}

We now show the feasibility  of the above proposed device for real application by analyzing
 the robustness of the antiresonance gap [i.e.,
the solid blue curve in Fig.\ \ref{fig2}(a)] against the Anderson
disorder.  The
converged conductance which is averaged over 3000 random configurations is plotted in
Fig.\ \ref{fig3}(a) against the Fermi energy in the vicinity of the
gap for different Anderson disorder strength $W$. From the
figure, one can see that the leakage conductance near
the gap is extremely small ($G<0.02e^2/h$)
until the strength of the disorder exceeds $0.04t$ which is three times
larger than the Zeeman splitting. For the large disorder strength $W=0.04t$, the corresponding
leakage current for the energy window
 $[0.0040t,0.0045t]$  is smaller than $1.5\times10^{-5}e^2t/h$,
more than one order of magnitude smaller than the ``on'' current
in the same energy interval. The leakage current is even much
  smaller for $W=0.01t$ ($0.02t$), i.e.,
$I\simeq7\times10^{-7}e^2t/h$ ($2\times10^{-6}e^2t/h$). For comparison, we also check the robustness of the
previous proposed transistors.\cite{feng,sanchez}
 In Fig.\ \ref{fig3}(b), the results of $T$-shaped
structure without the local Rashba SOC are plotted. The conductance
increases rapidly with the strength of the disorder, specifically, it
reaches $0.05e^2/h$ already at $W\sim0.02t$. Similar feature is also
obtained for the device with only the Fano antiresonance where the length of the side arm  $L_s=0$ and
$\lambda=0.13t$, as shown in Fig.\ \ref{fig3}(c). Therefore, the transistors based on the
structure antiresonance or the Fano antiresonance alone are very weak against the disorder and do not
provide an energy window, both in contrast to our new scheme which combine both the
Fano and the structure antiresonance together.
We also checked the robustness of the antiresonance gap against the 
disorder of the Rashba
 SOC, and obtained results very similar to the case with the on-site
  disorder. The average leakage conductance in the gap is about 
$0.01e^2/h$ with a disorder strength $W^\prime=0.5\lambda$, which is
 much smaller than the conductance with the same disorder strength at
the Fano-antiresonance point in structure without sidearm ($\sim 0.08e^2/h$). 

In summary, we have proposed a scheme for spin transistor by studying
a $T$-shaped structure with local Rashba SOC. Both leads are assumed to be half-metallic.
 The relevant conductance can be strongly
modulated by the Fermi energy of the leads, the strength of the Rashba SOC
 and the length of the sidearm. We have also demonstrated
that a wide antiresonance energy gap can be obtained by adjusting the Fano
antiresonance and the structure antiresonance close to each other. We propose
that our structure can be used as spin transistors, since a large current
can be obtained in the same antiresonance energy gap region when the two types of the antiresonance
are tuned away from each other by either change the Rashba coefficient and/or
change the length of the side arm electronically. We also show the robustness of
the antiresonance energy gap against the on-site disorder.
The wide working energy window (in contrast  to a single energy) and the much improved
 robustness against disorder suggest the proposed
structure has great potential for real application.

This work was supported by the Natural
Science Foundation of China under Grant Nos.\ 10574120 and 10725417, the
National Basic Research Program of China under Grant
No.\ 2006CB922005 and the Knowledge Innovation Project of Chinese Academy
of Sciences. One of the authors (MWW) was also partially supported by the
Robert-Bosch Stiftung and GRK 638. He would like to thank 
J. Fabian and C. Sch\"uller
at Universit\"at Regensburg for hospitality where this work
was finalized and J. Fabian for valuable
discussions. S.K. acknowledges J. Zhou for valuable discussions.

\end{document}